**Secondary scintillation yield of Xenon with sub-percent levels of $CO_2$ additive: efficiently reducing electron diffusion in HPXe optical TPCs for rare-event detection**

## The NEXT Collaboration

C.A.O. Henriques,[a] E.D.C. Freitas,[a] C.D.R. Azevedo,[c] D. González-Díaz,[i] R.D.P. Mano,[a] M.R. Jorge,[a] L.M.P. Fernandes,[a] C.M.B. Monteiro,[a,1] J.J. Gómez-Cadenas,[b,2] V. Álvarez,[b] J.M. Benlloch-Rodríguez,[b] F.I.G.M. Borges,[d] A. Botas,[b] S. Cárcel,[b] J.V. Carríon,[b] S. Cebrían,[e] C.A.N. Conde,[d] J. Díaz,[b] M. Diesburg,[f] R. Esteve,[g] R. Felkai,[b] P. Ferrario,[b] A.L. Ferreira,[c] A. Goldschmidt,[h] R.M. Gutiérrez,[j] J. Hauptman,[k] A.I. Hernandez,[j] J.A. Hernando Morata,[i] V. Herrero,[g] B.J.P. Jones,[l] L. Labarga,[m] A. Laing,[b] P. Lebrun,[f] I. Liubarsky,[b] N. López-March,[b] M. Losada,[j] J. Martín-Albo,[b,3] G. Martínez-Lema,[l] A. Martínez,[b] A.D. McDonald,[l] F. Monrabal,[l] F.J. Mora,[g] L.M. Moutinho,[c] J. Muñoz Vidal,[b] M. Musti,[b] M. Nebot-Guinot,[b] P. Novella,[b] D.R. Nygren,[l,1] B. Palmeiro,[b] A. Para,[f] J. Pérez,[b] M. Querol,[b] J. Renner,[b] L. Ripoll,[n] J. Rodríguez,[b] L. Rogers,[l] F.P. Santos,[d] J.M.F. dos Santos,[a] A. Simón,[b] C. Sofka,[o,4] M. Sorel,[b] T. Stiegler,[o] J.F. Toledo,[g] J. Torrent,[b] Z. Tsamalaidze,[p] J.F.C.A. Veloso,[c] R. Webb,[o] J.T. White,[o,5] N. Yahlali[b]

[a] *LIBPhys, Physics Department, University of Coimbra*
  *Rua Larga, 3004-516 Coimbra, Portugal*

[b] *Instituto de Física Corpuscular (IFIC), CSIC & Universitat de València*
  *Calle Catedrático José Beltrán, 2, 46980 Paterna, Valencia, Spain*

[c] *Institute of Nanostructures, Nanomodelling and Nanofabrication (i3N), Universidade de Aveiro*
  *Campus de Santiago, 3810-193 Aveiro, Portugal*

[d] *LIP, Departamento de Física, Universidade de Coimbra*
  *Rua Larga, 3004-516 Coimbra, Portugal*

[e] *Laboratorio de Física Nuclear y Astropartículas, Universidad de Zaragoza*
  *Calle Pedro Cerbuna, 12, 50009 Zaragoza, Spain*

[f] *Fermi National Accelerator Laboratory*
  *Batavia, Illinois 60510, USA*

[g] *Instituto de Instrumentación para Imagen Molecular (I3M), Universitat Politècnica de València*
  *Camino de Vera, s/n, Edificio 8B, 46022 Valencia, Spain*

[h] *Lawrence Berkeley National Laboratory (LBNL)*
  *1 Cyclotron Road, Berkeley, California 94720, USA*

[i] *Instituto Gallego de Física de Altas Energías, Univ. de Santiago de Compostela*
  *Campus sur, Rúa Xosé María Suárez Núñez, s/n, 15782 Santiago de Compostela, Spain*

[j] *Centro de Investigación en Ciencias Básicas y Aplicadas, Universidad Antonio Nariño*
  *Sede Circunvalar, Carretera 3 Este No. 47 A-15, Bogotá, Colombia*

[k] *Department of Physics and Astronomy, Iowa State University 12*

---

[1] Corresponding author. E-mail address: cristina@gian.fis.uc.pt
[2] NEXT Co-spokesperson.
[3] Now at University of Oxford, United Kingdom.
[4] Now at University of Texas at Austin, USA.
[5] Deceased.

Physics Hall, Ames, Iowa 50011-3160, USA

[l] Department of Physics, University of Texas at Arlington
Arlington, Texas 76019, USA

[m] Departamento de Física Teórica, Universidad Autónoma de Madrid
Campus de Cantoblanco, 28049 Madrid, Spain

[n] Escola Politècnica Superior, Universitat de Girona
Av. Montilivi, s/n, 17071 Girona, Spain

[o] Department of Physics and Astronomy, Texas A&M University
College Station, Texas 77843-4242, USA

[p] Joint Institute for Nuclear Research (JINR)
Joliot-Curie 6, 141980 Dubna, Russia

[1] Corresponding Author E-mail address: cristina@gian.fis.uc.pt

Abstract

We have measured the electroluminescence (EL) yield of Xe-$CO_2$ mixtures, with sub-percent $CO_2$ concentrations. We demonstrate that the EL production is still high in these mixtures, 70% and 35% relative to that produced in pure xenon, for $CO_2$ concentrations around 0.05% and 0.1%, respectively. The contribution of the statistical fluctuations in EL production to the energy resolution increases with increasing $CO_2$ concentration and, for our gas proportional scintillation counter, it is smaller than the contribution of the Fano factor for concentrations below 0.1% $CO_2$. Xe-$CO_2$ mixtures are important alternatives to pure xenon in Time Projection Chambers (TPC) based on EL signal amplification with applications in the important field of rare event detection such as directional dark matter, double electron capture and double beta decay detection. The addition of $CO_2$ to pure xenon at the level of 0.05-0.1% can reduce significantly the scale of electron diffusion from 10 mm/$\sqrt{m}$ to 2.5 mm/$\sqrt{m}$, with high impact on the High Pressure Xenon (HPXe) TPC discrimination efficiency of the events through pattern recognition of the topology of primary ionization trails.

1. Introduction

High-pressure xenon (HPXe) time projection chambers (TPCs) gained increasing importance during the last years. The application of HPXe TPCs to rare event detection such as double beta decay (DBD) and double electron capture (DEC), with or without neutrino emission, as well as directional dark matter (DDM) has been proposed or implemented [1-7]. The physics behind these experiments is of paramount importance in contemporary particle physics and cosmology.

When compared to liquid xenon and double phase xenon TPCs [8-14], detection in the gas phase offers some important advantages. While the event detection in liquid TPCs allows for compactness and self-shielding, some features may be essential for the above experiments to succeed. The impact of background depends strongly on the achieved energy resolution, which is much better for event detection in gas than in liquid. Furthermore, event interaction in the gas will allow for discrimination of the rare event topological signature, as demonstrated for DBD and DEC detection [15,16,5], in contrast to the interaction in liquid, where the extremely reduced dimensions of the primary ionization trail rules out any possible trail pattern recognition.

In particular, optical TPCs based on electroluminescence (EL) amplification of the primary ionization signal are the most competitive alternatives to those based on charge avalanche amplification. For the latter, the limited charge amplification at high pressure impacts the energy resolution, yielding at present a best value around 3% at 2.5 MeV for a 1kg-scale prototype based on micromegas [17], to be compared to 0.7% obtained for an EL amplification prototype of similar dimensions [18]. In addition, when compared to conventional electronic readout of the charge avalanche, EL optical readout through a photosensor has the advantage of mechanically and electrically decoupling the amplification region, rendering more immunity to electronic noise, radiofrequency pickup and high voltage issues.

Absolute values of the EL light yield have been measured in uniform electric fields [19-21] and in the modern micropatterned electron multipliers, as GEM, THGEM, MHSP and micromegas [22-24]. The statistical fluctuations in the scintillation produced in charge avalanches are dominated by the statistical fluctuations in the total number of electrons produced in the avalanche, since all the electrons contribute to scintillation production. On the other hand, the statistical fluctuations in the EL produced for uniform electric fields below the gas ionization threshold are negligible when compared to those associated with the primary ionization formation [25]. The latter situation is most important when event to background discrimination is also based on the energy deposited in the gas, as is the case of DEC and neutrinoless double beta decay, where the best achievable detector energy resolution is important for efficient background rejection.

Event discrimination based on the topological signature of the ionization trail is related to the low electron drift velocity of xenon and, mainly, to its large electron diffusion. The large electron diffusion is determined by the inefficient electron energy loss in elastic collisions with the xenon atoms, in particular in the range of reduced electric fields of few tens of V/cm/bar, used to drift the primary ionization cloud towards the signal amplification region. Diffusion hinders the finer details of the ionization trail, especially for large drift distances, and the topological signature of the events has a weaker effect [26].

The aforementioned problem can be mitigated by adding a molecular gas, like $CO_2$, $CH_4$ or $CF_4$, to pure xenon. With the addition of such molecules, new molecular degrees of freedom from vibrational and rotational states are made available for electron energy transfer in inelastic collisions. In this case, the energy distribution of the ionization electron cloud in the drift region tends to build up around the energy of the first vibrational level, typically at ~ 0.1 eV, even in the presence of minute concentrations of molecular additives.

Until recently, it was believed that the presence of molecular species in the noble gas would dramatically reduce the EL yield that could be achieved. Experimental studies performed for Ar [27] have shown that the presence of $CO_2$ and $CH_4$ in concentrations as low as ~ 50 ppm and ~ 200 ppm, respectively, resulted in an EL reduction above 70%. Detailed Monte Carlo simulation studies of electron drift in xenon at atmospheric pressure and room temperature [28] have shown that the average number of elastic collisions between successive inelastic collisions is very large, of the order of $10^4$, for the typical reduced electric fields applied to the scintillation region. This fact partly explained the importance of gas purity for the EL yield of noble gases: if an electron has significant probability of colliding with a molecular impurity before it obtains from the electric field sufficient energy to excite a noble gas atom, it may lose part of its energy without leading to EL photon emission, resulting in a decrease in the EL production. Depending on the conditions, excimer quenching, photo-absorption and dissociative attachment can jeopardize performance as well.

Within the NEXT collaboration [1], which has built a HPXe TPC for DBD studies with the $^{136}$Xe isotope, we proposed to revisit the addition of molecular additives to xenon, at sub-percent level, to reduce electron diffusion in the TPC, hence improving the topological discrimination capabilities. Preliminary experimental studies and simulations for different concentrations of $CO_2$ and $CH_4$ gases that are common in TPCs and whose elementary cross-sections are well known, have shown encouraging results [29], leading to acceptable EL losses and only small degradation in the detector energy resolution. Simulation results obtained with Magboltz have shown that Xe-$CO_2$ mixtures with concentrations of 0.05-0.1% of $CO_2$ would be sufficient to reduce the transversal and longitudinal diffusion coefficients to acceptable values (almost one order of magnitude lower than the concentrations needed for $CH_4$ to obtain a similar diffusion reduction [29]).

Those results led us to perform a detailed experimental study on the effect of the addition of $CO_2$ to pure xenon both on the EL yield and on the detector energy resolution, for different additive concentrations below 1% [29]. In this work we present those studies. $CO_2$ is a priori the most interesting option due to its low cost and easy handling, since it is non-flammable.

2. **Experimental setup**

The experimental setup especially projected for these studies includes a Gas Proportional Scintillation Counter (GPSC) [25], which is connected to a gas re-circulation system in order to continuously purify the gas or the mixture using SAES St-707 getters; a Residual Gas Analyser (RGA) that provides a real-time direct measurement of the molecular additive concentrations; a vacuum pumping system to maintain the RGA in continuous operation; associated electronics and suitable control and data-acquisition electronics for both systems, the RGA and the GPSC. The main components of the experimental setup are illustrated in Fig. 1.

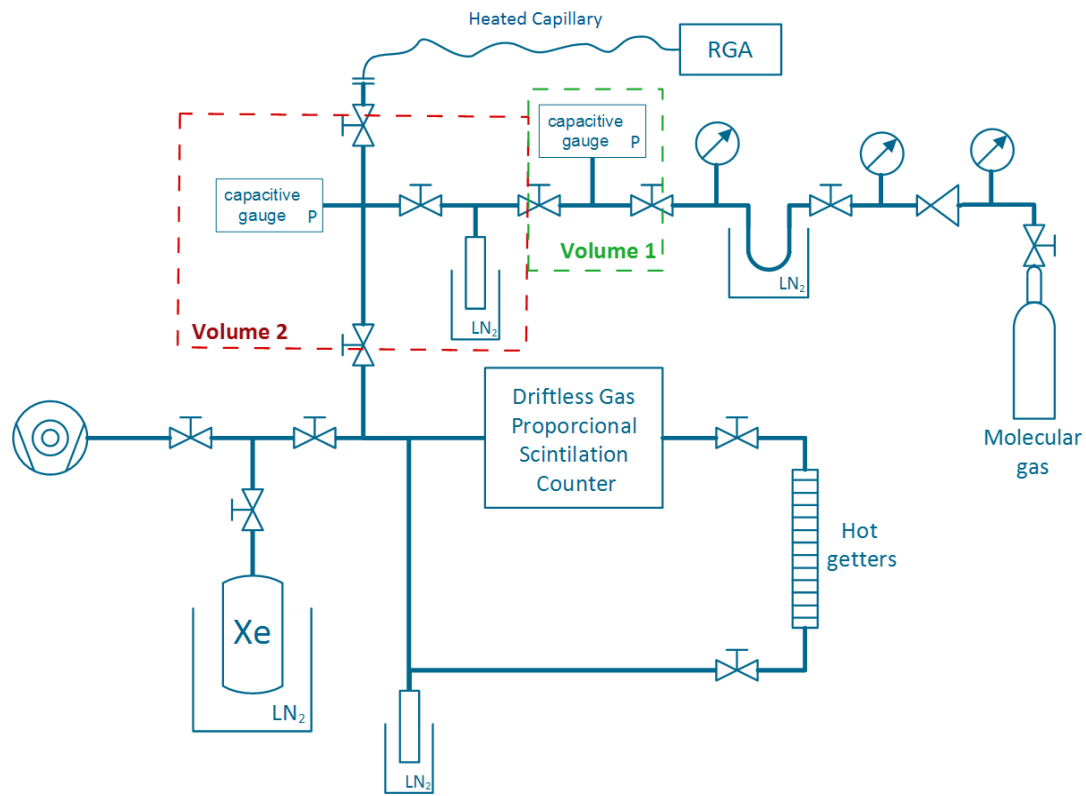

Fig. 1. Schematic of the experimental apparatus, including the gas proportional scintillation counter, the gas circulation and purifying system with SAES St707 getters, the gas entrance and exit systems including the turbo-pump, two calibration volumes (volume 1 in green, volume 2 in red), the liquid nitrogen mixing vessels and the RGA connection through a heated capillary.

The GPSC used in this work is depicted in Fig.2. It is of the 'driftless' type, i.e. without drift region, and has been already used in [30, 31]. This design was chosen for the present work because it allows to study the influence of molecular additives on the secondary scintillation parameters, minimizing the effects that may arise in the electron drift through the drift region and gas scintillation transparency.

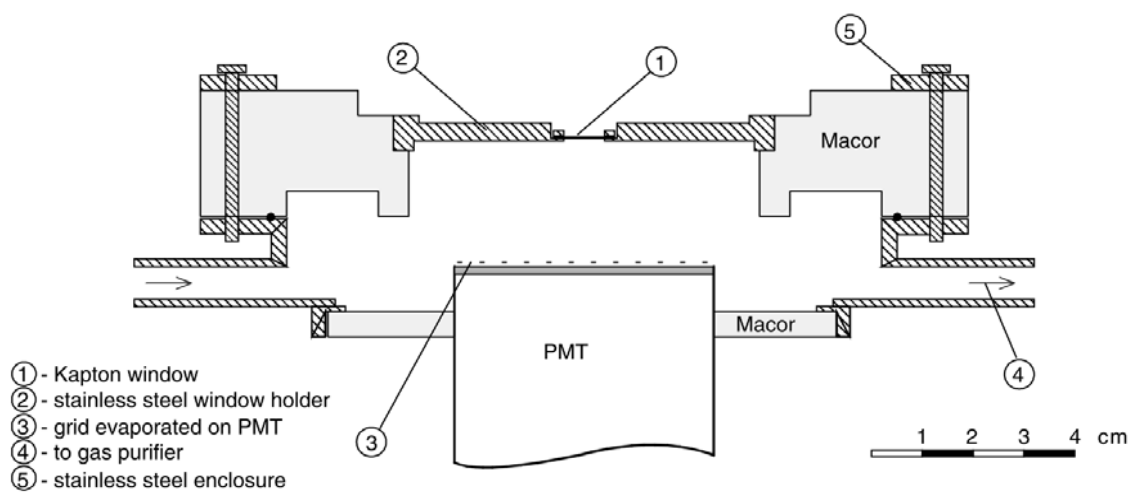

① - Kapton window
② - stainless steel window holder
③ - grid evaporated on PMT
④ - to gas purifier
⑤ - stainless steel enclosure

Fig. 2. Scheme of the driftless GPSC used in this work, including its principal components.

The EL region, 2.5 cm long, is delimited by a Kapton radiation window (8 mm in diameter, mounted on a stainless steel holder) aluminised on the inner side, and by the PMT quartz window, vacuum-evaporated with a chromium grid (100 µm width and 1000 µm spacing), electrically connected to the photocathode pin of the PMT. The EL electric field is established by applying a negative high voltage to the detector window and its holder, which are insulated from the stainless steel detector body by a ceramic material (Macor), being the detector body, the chromium grid of the PMT window and its photocathode kept at 0V. More information on this detector can be found in [30,31]. The reduced electric field inside the detector is set below the gas ionization threshold in order for EL to be produced without any charge multiplication in the scintillation region.

A 2-mm collimated, 5.9 keV x-ray beam from a $^{55}$Fe radioactive source irradiated the detector window along the detector axis, being the 6.4-keV x-rays of the Mn $K_\beta$ line absorbed by a chromium film. The 5.9 keV x-rays interact in the gas resulting in the release of electrons and photons. These ionization electrons are accelerated throughout the scintillation region exciting the noble gas atoms and inducing EL as a result of the atoms' de-excitation processes. The amount of EL is more than 3 orders of magnitude higher than primary scintillation. The EL pulse is collected by the PMT, whose output signal is subsequently shaped, amplified and, finally, digitized through a Multi-Channel Analyser (MCA). A typical pulse-height distribution obtained in the MCA for 5.9 keV x-rays is depicted in Fig.3a.

In a driftless chamber, the amount of EL scintillation depends on the distance travelled by the primary electron cloud in the scintillation region and, therefore, on the x-ray interaction depth. Consequently, the pulse-height distribution generated by the MCA has the typical Gaussian shape (from a monoenergetic line) convoluted with an exponential tail towards the low-energy region, due to the exponential law of the x-ray attenuation. Since, for 5.9-keV x-rays, the absorption length in 1 bar of xenon is about 2.7 mm, very small when compared to the long EL region of 25 mm, the observed full absorption peak in the pulse-height distribution has an almost Gaussian-like shape.

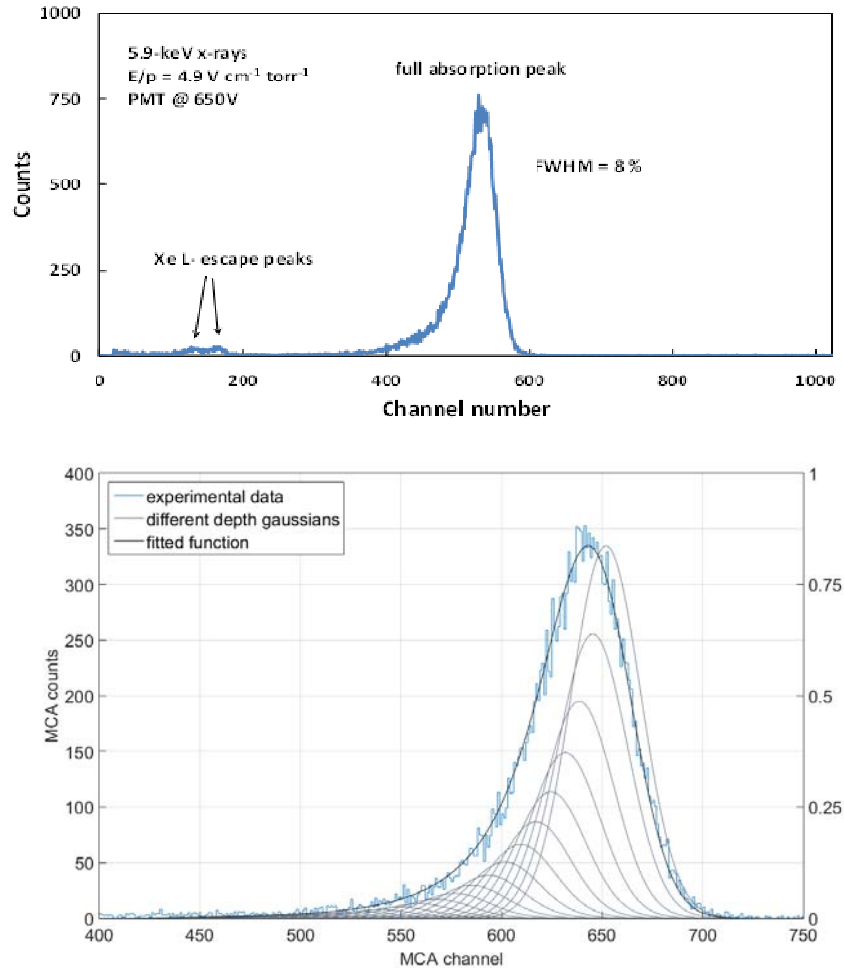

Fig.3. a) Pulse-height distribution for 5.9-keV x-rays absorbed in the xenon driftless GPSC for a reduced electric field of 3.7 kV/cm/bar; b) detail of the full energy peak together with the fit function resulting from the deconvolution procedure used in this work.

The intrinsic response of the GPSC for 5.9 keV x-rays was obtained by deconvolution of the overall full absorption peak distribution into a sum of a large number, 250, of Gaussian functions corresponding to x-ray interactions at successive depths, $\Delta z = 0.01$ cm, with areas decreasing according to the exponential absorption law for the 5.9 keV x-rays and with the same FWHM, which was left as a free parameter. Figure 3b) depicts an experimental pulse-height distribution for the full energy peak and the resulting fit obtained by the deconvolution procedure used, denoting a very good agreement. The GPSC pulse amplitude and energy resolution were taken from the centroid and FWHM of the Gaussian corresponding to X-ray interactions taking place just below the window. The obtained amplitude is within 2% of the peak channel of the full energy peak, Fig.3b. The obtained energy resolution is somewhat below 7%, instead of the ~8% obtained for a Gaussian fit to the right side.

The small volumes in fig. 1, each one read by an accurate capacitive pressure gauge, were used to calibrate the RGA, avoiding any error that might result from the $CO_2$ adsorption in the inner surfaces of the GPSC and, mainly, in the getters. The calibration process has shown a good linear correlation with the RGA analysis, within the studied concentration range. In order to avoid a pressure-dependent non-linearity of the RGA, calibration and detector operation have been carried out at the same total pressure of about 1.13 bar, for both pure xenon and its mixtures. The EL studies were performed when the RGA partial pressures stabilized and, likewise, the additive concentration was calculated from an average over several measurements done during the time interval when the EL studies were performed.

Before setting each mixture, a measurement of the $CO_2$ background was performed in the GPSC filled with pure xenon, being the getters at 250 °C, in order to ensure maximum xenon purity. This background was, afterwards, subtracted from the RGA $CO_2$ reading once the mixture had been prepared. As the hot getters react with $CO_2$, the getter temperature was reduced to 80 °C at this point. At this getter temperature $CO_2$ is only slightly absorbed. On the other hand, the EL parameters in pure xenon were found to degrade only slightly and only after several days of operation when cooling down the getters from 250 °C to 80 °C. Furthermore, it was observed that, despite the $CO_2$ being absorbed in the getters, part of it is also transformed into CO that escapes to the gas phase, as observed by the correlated growth of the partial pressure at mass 28 (related with CO) as the concentration of $CO_2$ decreases. This effect increases with increasing getter temperature. Therefore, some CO is present in the Xe-$CO_2$ mixtures, with concentrations that are roughly constant for all the studied mixtures, for a getter operation temperature of 80 °C. Simulations have shown that the impact of the presence of CO on the yield is small, within a 10% effect at most, for the lowest $CO_2$ concentrations.

## 3. Experimental results

In order to obtain absolute values for the EL reduced yield, $Y/N$, $N$ being the density of the molecules in the gas, the relative variation of the pulse amplitude as a function of the reduced electric field, $E/N$, obtained for pure xenon, was normalized to the experimental EL reduced yield obtained in [19]. The same normalization constant was, then, used to normalize the remaining EL yield curves obtained for the different mixtures.

In Fig. 4, $Y/N$ as a function of the $E/N$ applied to the scintillation region is shown for different $CO_2$ concentrations added to pure xenon. The two data sets for the 0.174% of $CO_2$ are related to two independent measurements. Interestingly, the reduced yield exhibits the typical approximate linear dependence of EL with reduced electric field even in the presence of $CO_2$. The solid lines present fits to the data, excluding the data points near the EL threshold where the EL response of GPSCs deviates from the linear trend [30]. As expected, the EL yield decreases as the $CO_2$ amount increases. Nevertheless, $CO_2$ concentrations of ~0.05%, which allow an overall electron diffusion around 2.5 mm/$\sqrt{m}$ [29], can be acceptable in terms of EL yield since, in spite of having an EL yield reduction up to 35% when compared to pure xenon, this reduction may be tolerable in the cases where the EL scintillation is large enough. For comparison, it must be recalled that such a reduction is achieved in Ar mixtures when $CO_2$ concentration reaches 10 ppm [27].

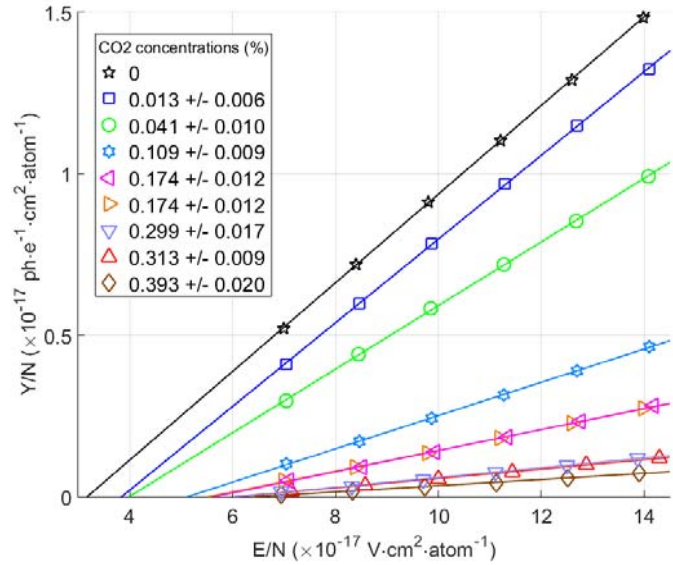

Fig. 4. EL reduced yield, Y/N, as a function of the reduced electric field, E/N, for different concentrations of $CO_2$ added to pure xenon. The solid lines are linear fits to the data.

As anticipated, for the same reduced electric field intensity, the EL scintillation threshold increases with increasing CO2 content since, upon colliding with a CO2 molecule, the electron loses energy to rotational and vibrational excited states, reducing the average electron energy. Although qualitative in nature, the behaviour of the scintillation threshold shows how this cooling seems to be very efficient up to concentrations around 0.1% (as indicated by Magboltz simulations), values for which the scintillation loss remains acceptable, hinting that a compromise in terms of electron cooling/excimer scintillation does exist. Additional losses can be expected in $CO_2$ from dissociative attachment and excimer quenching, this last one being indeed the main source identified earlier in [27]. Figure 5 summarizes the EL scintillation threshold and reduced yield slope, from Fig.4 data, as a function of $CO_2$ concentration.

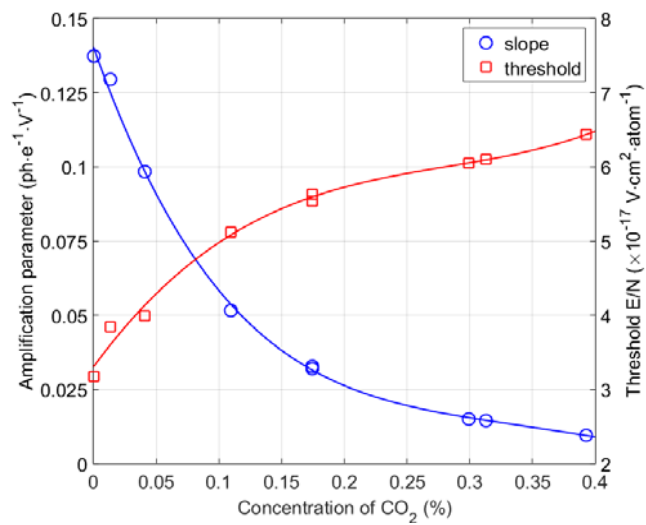

Fig. 5. Light amplification parameter and EL threshold of the lines fitted to the reduced EL yield (Fig. 4). The solid lines serve only to guide the eye.

The impact of the molecular additive in the TPC energy resolution is an important parameter to be considered, in particular in double electron capture and in neutrinoless double beta decay detection, as it is a tool to effectively discriminate the rare events against background. In Fig. 6 we present the GPSC energy resolution for 5.9 keV x-rays as a function of the reduced electric field, for the different $CO_2$ concentrations used in this work.

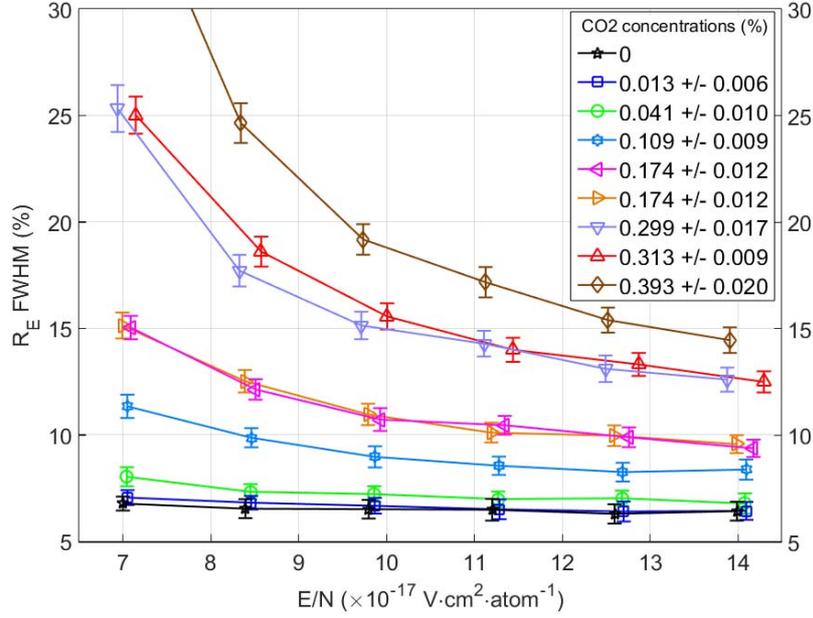

Fig. 6. Detector energy resolution as a function of E/N for 5.9 keV x-rays and for different concentrations of $CO_2$. The solid lines serve only to guide the eye.

The energy resolution of a GPSC, $R_E$, is given by [25]:

$$R_E = 2\sqrt{2 \ln 2} \sqrt{\frac{\sigma_e^2}{\bar{N}_e^2} + \frac{1}{\bar{N}_e}\left(\frac{\sigma_{EL}^2}{\bar{N}_{EL}^2}\right) + \frac{\sigma_{pe}^2}{\bar{N}_{pe}^2} + \frac{1}{\bar{N}_{pe}}\left(\frac{\sigma_q^2}{\bar{G}_q^2}\right)} \quad \text{(Eq.1)}$$

where the first term under the square root describes the relative fluctuations in the number of ionization electrons induced by the interaction, $N_e$, the second term describes the relative fluctuations associated to the number of EL photons produced in the scintillation region per primary electron, $N_{EL}$, and the last two terms describe the relative fluctuations in the photosensor, namely the relative fluctuations in the number of photoelectrons released from the PMT photocathode by the EL burst, $N_{pe}$, and the relative fluctuations in the number of electrons collected in the PMT anode per photoelectron, i.e. the relative fluctuations in the gain of the electron avalanche in the PMT dynodes. Since the process of photoelectron release from a photocathode by the incoming photons is described by a Poisson distribution, its variance is $\sigma_{pe}^2 = \bar{N}_{pe}$ and the relative fluctuations in the PMT are given by:

$$\frac{1+\left(\frac{\sigma_q^2}{\bar{G}_q^2}\right)}{c\bar{N}_e\bar{N}_{EL}} = \frac{k}{\bar{N}_e\bar{N}_{EL}} \quad \text{(Eq.2)},$$

c represents the light collection efficiency, related to the anode grid transparency (Fig. 2), the PMT quantum efficiency and the average solid angle subtended by the PMT photocathode relative to the primary electron path in the EL region. Therefore, $k$ is a constant, which depends on the scintillation readout geometry and on the photosensor itself.

From the data for pure xenon, we can experimentally determine the contributions to the energy resolution from the statistical fluctuations due to the primary ionization formation and due to the photosensor, since the contribution from the statistical fluctuations to the EL is negligible when compared to the other factors [25]. In Fig. 7 we depict $R_E^2$ as a function of $N_{EL}^{-1}$ for pure xenon. A linear function, as imposed by Eq.1, is fitted to the data points, excluding those with the highest and the lowest $N_{EL}$, which depart from the linear trend. This behaviour is similar to the one observed in standard type GPSCs, with drift region [32,33]. The first term is obtained from the line interception with the vertical axis while $k$ is given by the slope of the line. From Fig.7 we obtain a Fano factor $F = \sigma_e^2/\bar{N}_e$ = 0.17 +/- 0.04 for xenon, using a $w$-value of 22 eV, i.e. the average energy needed to produce one electron-ion pair in xenon ($N_e = E_x/w$, $E_x$ being the x-ray energy). This result is in good agreement with the values normally found in the literature, between 0.13 and 0.25 [32, 34-36]. In addition, the result we obtain for $k$ is similar to what is obtained from calculations based on the geometry and the PMT characteristics of our setup. These agreements show the robustness of the present method.

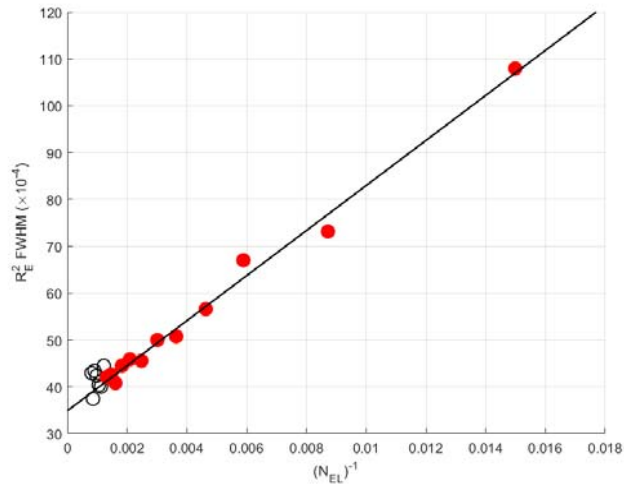

Fig. 7. The square of the energy resolution ($R_E$) as a function of the inverse of the average number of photons produced in the scintillation region per primary electron for 5.9 keV x-rays and for pure xenon. The solid line presents a linear fit to the data points in red.

The terms of Eq. 1 obtained from the fit in Fig.7 (the 1st and the last 2), for the relative fluctuations in the primary ionization formation and for the relative fluctuations in the photosensor, are constant for all $CO_2$ concentrations studied in this work, since the Fano factor and the $w$-values of those mixtures are not expected to change significantly, due to the fact that the additive concentration is very small and $k$ is constant for a given geometry and photosensor setup. Therefore, it is possible to determine the fluctuations associated to the EL production as a function of reduced electric field for the different mixtures, using the $R_E$ data from Fig.6, as these fluctuations are the only unknown variable in Eq.1.

In Fig. 8 we present the square of the relative standard deviation in the number of EL photons produced in the scintillation region per primary electron,

$$Q = \left(\frac{\sigma_{EL}^2}{\overline{N}_{EL}^2}\right) \quad \text{(Eq.3)},$$

as a function of reduced electric field in the scintillation region, for the different concentrations of $CO_2$ added to pure xenon. A striking observation that can be made in Fig.8 is that $Q$ becomes non-negligible as the $CO_2$ concentration increases, largely independent of the reduced electric field. For a $CO_2$ concentration as low as 0.1% $Q$ ~ 0.08, i.e. about half the value of the Fano factor, while for 0.2% $CO_2$ $Q$ becomes comparable. For a CO2 concentration of 0.05% $Q$ is found to be around 0.02, a value that has a negligible impact on the energy resolution. For higher CO2 concentrations and lower E/N, the signal-to-noise ratio decreases significantly, resulting in an artificially high energy resolution and, consequently, an over-estimated Q value obtained from Eq.1, as the noise is not included in this equation. For that reason, those points are not included in Fig.8.

The rise in the contribution from $Q$ cannot be explained if we take only into account the effect of EL reduction with increasing $CO_2$, since even a reduction of one order of magnitude in EL still means a very high number of EL photons and, in addition, one would expect a decrease in $Q$ with increasing electric field in the scintillation region, instead of an almost constant trend. We believe that this effect is due to dissociative electron attachment to $CO_2$ molecules, for which we find a good quantitative agreement with simulations [37]; in the absence of this effect, simulations predict values for the Q-factor around 0.02, little dependent on concentration. Indeed, the cross-section for electron attachment is non-negligible and has narrow peaks for electron energies between 4 - 5 and 7 - 9 eV, e.g. see Fig.1 in [38], and these are energies that the electrons will eventually reach in the scintillation region in order to be able to excite the xenon atoms. The simulated attachment implies for the highest $CO_2$ concentration a relatively modest 10's of %-level loss of ionization electrons, an effect which does not dominate the EL production. Its presence becomes nonetheless the main source of fluctuations in the EL signal for concentrations already above 0.17% $CO_2$.

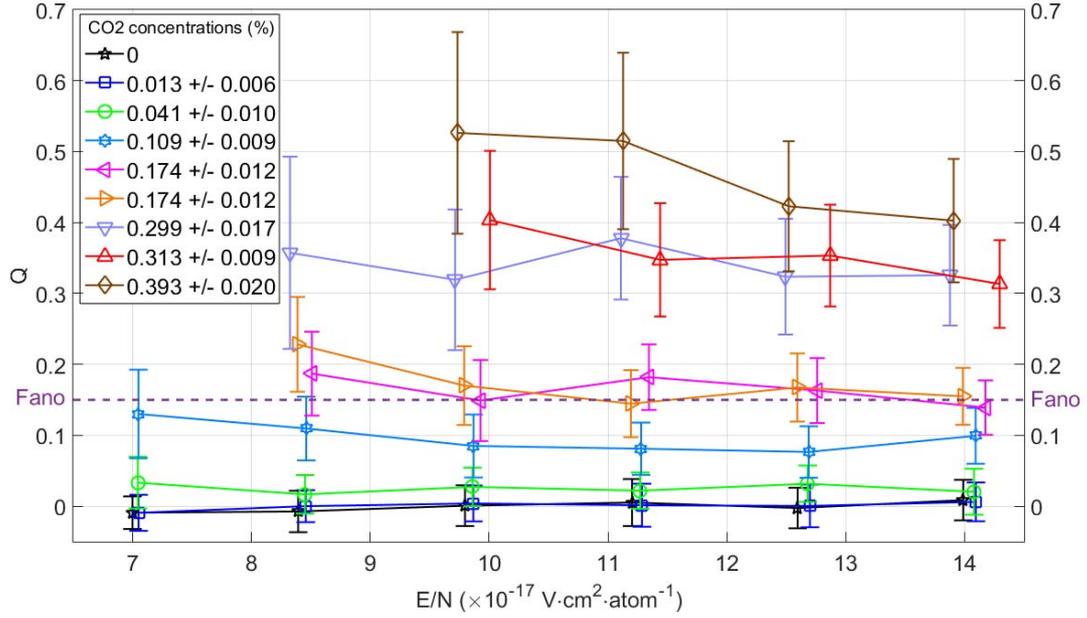

Fig. 8. Relative standard deviation in the number of EL photons per primary electron once squared (Q), as a function of the reduced electric field in the scintillation region (E/N) for different concentrations of $CO_2$. The solid lines serve only to guide the eye.

## 4. Discussion

$CO_2$ concentrations in the range of 0.05%-0.1% correspond to a characteristic size of the electron diffusion ellipsoid of $\sqrt[3]{\sigma_x \sigma_y \sigma_z} \cong 2.5\ mm\ \times \sqrt{\frac{10\ bar}{P}}$ after a 1m drift through the TPC [29,37]. This value can be found for reduced drift fields in the range of E/P = [20-30] V/cm/bar, by resorting to the latest Magboltz cross-section database. The drastic change experimentally observed in the scintillation threshold (Fig.5) suggests that electron cooling is in fact strongly active even for these minute concentrations. Moreover, simulations indicate that a minimum of diffusion exists in the above field range; therefore, there is little to be gained by increasing (or decreasing) the drift field in the TPC. The presence of such minima can be found experimentally and theoretically for xenon mixtures and additives like $CH_4$ or TMA in [17], and it becomes narrower at low concentrations.

In pure xenon at 10 bar, electric fields in the aforementioned range pose no problems concerning charge recombination for primary electrons, as can be readily noticed, for instance, in [24]. For admixtures, however, the situation at high pressures is less clear. Some qualitative arguments can be drawn: in xenon at 5 bar and in a 0.22% TMA admixture, for example, the additional contribution to the Fano factor stemming from fluctuations in the charge recombination process is less than 0.1 [39]. Since the measured diffusion and drift coefficients are, in that case, similar to those simulated for the $CO_2$ optimum (0.05%-0.1%), and being the ionization density close to the one attempted in NEXT (10bar), measurements performed in Xe-TMA can be used to estimate an upper bound to the effect expected in Xe- $CO_2$. Besides this initial charge recombination, it must be noted that the drift velocity will be reduced for optimum $CO_2$ concentrations by a factor of around x2, which is not expected to have a dramatic effect on the electron lifetime, according to the measurements performed in [17], again for TMA admixtures.

The above aspects will soon be evaluated for high pressures in NEXT-DEMO, a large prototype with a drift length of 30 cm and a hexagonal cross section of 8-cm apothem, operated with ~ 1.5 kg of natural xenon at a pressure of 10 bar [40]. Other relevant effects like the pressure-dependence of the scintillation yields and the light fluctuations will also be evaluated. Simulations performed in [37] indicate that, despite the anticipated deterioration at high pressure, both the $Q$ factor and the finite-statistic term from the PMTs can be kept at the level of the Fano factor at 10 bar, as long as the $CO_2$ concentration remains in the range of 0.05-0.1%. Concerning the primary scintillation yields, a tolerable reduction within a factor of 5-10 is expected in the same concentration range.

A possible drawback that can arise from the use of $CO_2$ is related to the gas stability in the long term and the associated formation of CO. This can be handled using specific getters for $CO_2$. On the other hand, being these devices cold getters, one has to evaluate the radon emanation.

## 5. Conclusions

We have demonstrated that the addition of $CO_2$ to pure xenon, at concentration levels of few tenths of a percent, does not kill the proportional electroluminescence (EL) yield entirely, as it has been assumed during the last decades. $CO_2$ concentrations of 0.05% and 0.1% at around atmospheric pressure lead only to an EL reduction of 35% and 70%, respectively, relative to that produced in pure xenon at the same reduced electric field. Such a modest reduction seems tolerable, provided the number of photons produced per ionization electron is very large and also because it may be readily compensated by increasing the reduced electric field, since higher field can be applied to the scintillation region, as the ionization threshold increases with increasing $CO_2$ concentration.

On the other hand, the intrinsic energy resolution of xenon-based TPCs (i.e., photo-detection statistics neglected) degrades with increasing $CO_2$ concentration; for a concentration of 0.05% the contribution of the statistical fluctuations associated to EL production is a factor of 6 lower than the Fano factor, for 0.1% nearly half and for 0.2% it is slightly above it. This degradation in the energy resolution cannot be, however, compensated by an increase in the reduced electric field intensity. Based on both the approximate linear dependence of Q on the $CO_2$ concentration and the comparison with Magboltz simulations, these large fluctuations can be attributed to dissociative attachment of ionization electrons to CO2 molecules. Seemingly, this process can only be mitigated by using shallower EL regions. Nevertheless, a compromise has to be found between the thickness of this region and the amount of EL produced.

The above findings can be important for xenon-based TPCs relying on EL signal amplification, which are being increasingly used for rare-event detection such as directional dark matter, double electron capture and double beta decay detection. Particularly, the addition of $CO_2$ to pure xenon at the level of 0.05%- 0.1% will reduce significantly the electron transverse diffusion from 10mm/$\sqrt{m}$ to the level of few mm/$\sqrt{m}$, having a high impact on the discrimination of events through pattern recognition of the topology of primary ionization trails.

Other molecular additives, such as $CH_4$, do not present the drawback of having significant electron attachment but, on the other hand, higher concentrations will be needed to obtain similar electron transverse and longitudinal diffusions as in $CO_2$. Nevertheless, former work in [27] has shown that the addition of $CH_4$ to pure argon has less impact on the reduction in the mixture scintillation yield when compared to the addition of $CO_2$.

**Acknowledgements**

The NEXT Collaboration acknowledges support from the following agencies and institutions: the European Research Council (ERC) under the Advanced Grant 339787-NEXT; the Ministerio de Economía y Competitividad of Spain under grants FIS2014-53371-C04 and the Severo Ochoa Program SEV-2014-0398; the GVA of Spain under grant PROMETEO/2016/120; the Portuguese FCT under project PTDC/FIS-NUC/2525/2014; the U.S. Department of Energy under contracts number DE-AC02-07CH11359 (Fermi National Accelerator Laboratory) and DE-FG02-13ER42020 (Texas A&M); the University of Texas at Arlington. C.A.O.H., E.D.C.F., C.M.B.M. acknowledge FCT under grants PD/BD/105921/2014, SFRH/BPD/109180/2015, SFRH/BPD/76842/2011.